\documentclass[APS, Times1COL]{WileyNJDv5}

\articletype{Original Article}

\received{00}
\revised{00}
\accepted{00}
\usepackage{siunitx}

\begin{document}

\title{Quantifying Phase Magnitudes of Open-Source Focused-Probe 4D-STEM Ptychography Reconstructions}

\author[1]{Toma Susi}

\authormark{Toma Susi}
\titlemark{Quantifying Ptychographic Phase Magnitudes}

\address[1]{\orgname{University of Vienna}, \orgdiv{Faculty of Physics},
	\orgaddress{Boltzmanngasse 5, 1090 Vienna, \country{Austria}}}

\corres{toma.susi@univie.ac.at}

\abstract[Abstract]{Accurate computational ptychographic phase reconstructions are enabled
by fast direct-electron cameras with high dynamic ranges used for four-dimensional scanning 
transmission electron microscopy (4D-STEM). The availability of open software packages is 
making such analyses widely accessible, and especially when implemented in Python, easy to
compare in terms of computational efficiency and reconstruction quality. In this contribution, I 
reconstruct atomic phase shifts from convergent-beam electron diffraction maps of pristine 
monolayer graphene, which is an ideal dose-robust uniform phase object, acquired on a Dectris 
ARINA detector installed in a Nion UltraSTEM 100 operated at 60 keV with a focused-probe 
convergence semi-angle of 34 mrad. For two different recorded maximum scattering angle 
settings, I compare a range of direct and iterative open-source phase reconstruction algorithms, 
evaluating their computational efficiency and tolerance to reciprocal-space binning and
real-space thinning of the data. The quality of the phase images is assessed by quantifying the 
variation of atomic phase shifts using a robust parameter-based method, revealing an overall 
agreement with some notable differences in the absolute magnitudes and the variation of the 
phases. Although such variation is not a major issue when analyzing data with many identical 
atoms, it does put limits on what level of precision can be relied upon for unique sites such as 
defects or dopants, which also tend to be more dose-sensitive. Overall, these findings and the 
accompanying open data and code provide useful guidance for the sampling required for desired 
levels of phase precision, and suggest particular care is required when relying on electron 
ptychography for quantitative analyses of atomic-scale electromagnetic properties.}
	

\maketitle

\section{Introduction}

Aberration-corrected scanning transmission electron microscopy (STEM) is
a powerful tool for imaging~\cite{krivanek_atomatom_2010} and even
manipulation~\cite{susi_identifying_2022} of the atomic structure of materials, as
well as the characterization of their electronic and
vibrational~\cite{hage_singleatom_2020} properties with electron energy-loss
spectroscopy. Its success has been spurred by rapid developments
in instrumentation~\cite{hawkes_aberration_2009}, understanding of irradiation
damage~\cite{susi_quantifying_2019}, and advanced computational tools including
machine learning~\cite{kalinin_machine_2022}. Recently, analyzing all scattered
electrons by recording 2D diffraction patterns from 2D scanned areas in
a technique called 4D-STEM has become increasingly powerful for virtual
diffraction imaging, simultaneous resolving of light and heavy
elements~\cite{yucelen_phase_2018}, and mapping of phase, orientation and
strain as well as sample thickness and tilt~\cite{ophus_fourdimensional_2019}.~This is
increasingly practical due to the commercialization of fast and
sensitive direct-electron detectors~\cite{tate_high_2016,zambon_highframe_2023}, which enable not
only more accurate determination of the deflections of the electron
probe \cite{muller-caspary_measurement_2017}, but also access to the redistribution of
intensity within Bragg disks containing phase information.

Harnessing the redundancies in overlapping 4D datasets enables an
efficient means of scanning coherent diffractive imaging called
ptychography~\cite{hegerl_dynamische_1970}, where the complex electron probe and
sample potentials can be reconstructed for post-acquisition aberration
correction~\cite{yang_simultaneous_2016} and super-resolution
imaging~\cite{nellist_resolution_1995}, allowing sub-\AA\ projected spacings to be
precisely measured in 2D heterostructures~\cite{jiang_electron_2018}. Thin
specimen act as phase objects, causing phase shifts of the electron
waves that are directly correlated with atomic-scale electromagnetic
potentials. This has enabled the direct imaging of electrostatic potentials and the
charge density~\cite{murthy_spatial_2021,martis_imaging_2023,susana_direct_2024} as well as (at least) antiferromagnetic
order~\cite{kohno_realspace_2022,nguyen_angstromscale_2023,cui_antiferromagnetic_2024}. However, as the nuclei dominate electron
scattering, it is extremely challenging to reliably tease out valence
properties whose contribution is 10--100 times smaller, let alone the
even weaker magnetic ones. Simulations based on first-principles
scattering potentials are vital~\cite{madsen_initio_2022} to assist in the
measurement of charge transfer~\cite{martinez_direct_2023,hofer_detecting_2024}, but understanding
optimal sampling conditions~\cite{yang_efficient_2015,oleary_contrast_2021}~and differences between
algorithms~\cite{clark_effect_2023}, as well as quantifying the precision and
accuracy of phase reconstructions, are pressing issues that have not
been fully addressed yet.

Ptychographic reconstruction algorithms fall into two categories:
non-iterative direct~\cite{bates_subangstrom_1989} methods perform a reconstruction of
the full scan area, usually via Wigner-distribution deconvolution
(WDD)~\cite{rodenburg_theory_1992} or single-sideband ptychography
(SSB)~\cite{pennycook_efficient_2015}, while iterative ones refine it over many
iterations, most notably the ptychographic iterative
engine~\cite{rodenburg_phase_2004} and its later extensions~\cite{maiden_improved_2009}
and improvements~\cite{maiden_further_2017}. The latter also include
generalized maximum-likelihood methods that are not only robust with
respect to noise and probe aberrations, but also able to correct for
errors due to scan positioning and partial
coherence~\cite{thibault_reconstructing_2013}. Iterative methods with defocused
illumination allow large areas to be reconstructed at reduced
dose~\cite{chen_mixedstate_2020}, but using a focused probe is convenient as it
allows simultaneous atomically resolved imaging that helps elemental
identification via annular
dark-field \emph{Z}-contrast~\cite{pennycook_efficient_2015}. For more information 
on these algorithms, I refer the reader to recent literature~\cite{oleary_contrast_2021,varnavides_iterative_2024}.

Here, I analyze convergent-beam electron diffraction
maps recorded with an aberration-corrected focused probe with a
convergence semi-angle of 34 mrad. The sample is pristine one-atom thick
monolayer graphene, which represents an ideal weak-phase object where
each atom is equivalent and that is impervious to irradiation damage in
ultra-high vacuum at the 60~keV primary beam energy~\cite{susi_isotope_2016}.
The data were acquired on a retractable Dectris ARINA hybrid-pixel
detector~\cite{zambon_highframe_2023} installed on-axis in a Nion UltraSTEM 100
instrument, whose ultra-stable sample stage and flexible electron optics
make it ideally suited for 4D-STEM. The high quantum efficiency and dynamic
range of the detector, the irradiation stability of the specimen, the
low drift of the stage, and the high beam current of up to 200 pA together
ensure that these measurements constitute a nearly ideal dataset for benchmarking.

Using these data, I compare a number of open-source implementations of direct 
and iterative phase reconstruction algorithms: SSB and WDD, as well as integrated center of mass (iCOM; the equivalent method is dubbed iterative differential phase contrast in py4DSTEM)~\cite{savitzky_py4dstem_2021}, parallax-corrected bright-field imaging
(ie. tilt-corrected bright-field STEM~\cite{spoth_doseefficient_2017}), and batched
iterative gradient descent single-slice ptychography~\cite{varnavides_iterative_2024}
(which is essentially equivalent to the widely used ePIE~\cite{maiden_improved_2009} method, apart from the batching of probe positions which greatly improves convergence~\cite{varnavides_iterative_2024}). 
Computational times and the robustness of the reconstruction vary greatly depending 
on the algorithm and the sampling in both real and reciprocal space, although it should be also
acknowledged that different algorithms may be more or less sensitive to small discrepancies 
of input parameters that are unavoidable in practice. 

In the case of the present data, even with high 
sampling and areal doses of over 10\textsuperscript{6}~electrons/\AA\textsuperscript{2},
phase images show minor variations from the expected uniform atom contrast, 
partly due to imperfectly corrected residual aberrations, as well as differing sensitivities
to reciprocal-space binning and real-space sampling, as quantified using 
a parameter-based iterative method~\cite{hofer_reliable_2023}. These findings,
which include fully open access to both the data and the analysis code, thus provide
a useful resource for understanding and further developing electron-ptychographic
methods for quantitive phase imaging.

\section{Results}

\begin{figure}[b!]
	\begin{center}
		\includegraphics[width=0.9\columnwidth]{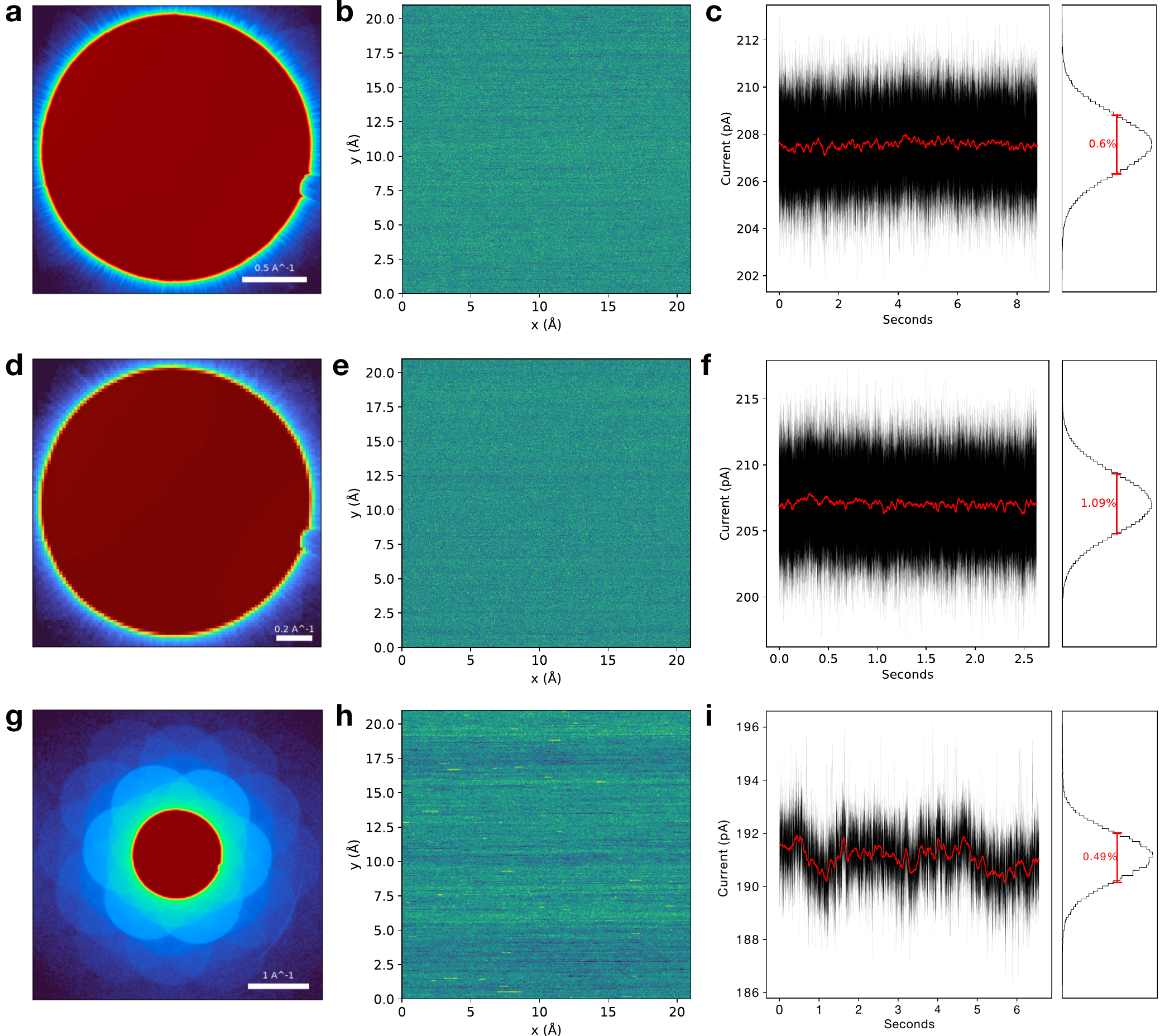}
		\caption{a) Mean diffraction pattern of the magnified projecter lens setting 
				(maximum scattering angle of 36 mrad) unbinned data 
				on a logarithmic intensity scale. b) Variation of the dose over the
				512$\times$512 scan pixels (pixel dwell time 33.3~\si{\us}). c)
				Variation of the current over time. The red curve is a moving average.
				(d-f) The same plots for the data with 2$\times$ 
				binning and 512$\times$512 scan pixels (pixel dwell time 10.0~\si{\us}).
				(g-i) The same plots for the extended projector lens setting 
				(maximum scattering angle of 109 mrad) with 256$\times$256
				scan pixels (pixel dwell time 100~\si{\us}).
				{\label{fig:cbed_and_dose}
				}%
		}
	\end{center}
\end{figure}

\subsection{Characteristics of the data}

Preparation of the pristine monolayer graphene samples, 4D-STEM data
acquisition, and phase reconstruction and quantification are described
in the Methods section, with all data and analysis code openly available 
(see Data and Code). I recorded datasets with two different maximum 
collected scattering angles controlled by the projector lens settings: a 
magnified one with 512$\times$512 real-space~\textbf{\emph{R}} scan 
positions where the maximum scattering angle in the vertical direction was 
36~mrad, and thus the bright-field disk almost covered the camera, and an 
extended one with a maximum scattering angle of 109~mrad, with 256$\times$256 
scan positions. For the magnified setting, I recorded data with the full
unbinned sensor with 192$\times$192 reciprocal-space~\textbf{\emph{Q}} pixels
(with the bright-field disk diameter of $\sim$177 pixels), where the data rate of the
10G fiber-optic connection to the camera limits pixel dwell time to 33.3~\si{\us},
and a 2$\times$ hardware-binned mode with 96$\times$96 \textbf{\emph{Q}} pixels
(bright-field disk of $\sim$89 pixels) that allows the fastest possible dwell time of 
10~\si{\us} per pixel (ie. 120 000 frames per second).
For the extended setting, because of the scattering cross section decreasing as
a function of scattering angle, I increased the pixel dwell time to 100~\si{\us}, and
recorded the full unbinned sensor to ensure good \textbf{\emph{Q}} sampling
(bright-field disk of $\sim$59 pixels).

The mean convergent-beam electron diffraction (CBED) pattern recorded
over graphene are shown in Figure~{\ref{fig:cbed_and_dose}}, alongside the spatial 
and time variation of the dose over the respective 8.8, 2.6, and 6.6~s acquisitions.
The standard deviation of recorded current was in the range of 1\% as expected for 
the cold-field emission gun, with the greatest time variation for the fastest pixel dwell 
time and with some variation visible in the slow-scan direction in the slow-scan 
direction across the horizontal horizontal scan lines. The irradiation doses were
counted directly from the diffraction patterns (thus ignoring the
negligible scattering outside the detector from such a weakly scattering
object) with a correction for the measured detective quantum efficiency
of the camera of 0.80 at 60~keV~\cite{zambon_highframe_2023}. The corresponding
beam currents for the two datasets were 207, 207 and 192~pA, resulting in
respective total doses of 11.4$\times$10\textsuperscript{9}, 3.4$\times$10\textsuperscript{9},
and 7.8$\times$10\textsuperscript{9} electrons and areal doses of
2.6$\times$10\textsuperscript{7}, 0.77$\times$10\textsuperscript{7}, and 
1.77$\times$10\textsuperscript{7}    
electrons/\AA\textsuperscript{2}. These doses are notably larger than typical works on
focused-probe low-dose reconstructions~\cite{oleary_phase_2020,hofer_reliable_2023, 
ooe_doseefficient_2024}, 
but as single-layer graphene is a weakly scattering object, direct comparison of 
fluences might be misleading. However, comparing the two datasets with different 
pixel dwell times allows us to assess this directly.

\subsection{Phase reconstructions}

\subsubsection{Magnified projector lens setting: software binning in~\textbf{\emph{Q}}}
First, I compare phase images reconstructed from the 
full dataset with 512$\times$512 real-space~\textbf{\emph{R}} scan positions 
with the magnified projector lens setting (36 mrad maximum scattering angle)
with different degrees of reciprocal-space~\textbf{\emph{Q}} binning. Notably, the
parallax, SBB and WDD algorithms could not cope with the memory requirements
of the full $\sim$39~GB dataset on our processing workstation with
128~GB of memory (see Methods), and I will therefore compare binning factors of 2 and above.
The uncompressed data array sizes alongside reconstruction times for different algorithms are 
shown in Table~{\ref{tab:data}}. In terms of computational
timings on this hardware, all algorithms perform significantly faster when data is binned
in~\emph{\textbf{Q}}, although there seems to be little further benefit
from increasing the bin factor from 16 to 32. Comparing the speed of
the algorithms is caveat by the fact that I used greater~\textbf{\emph{Q}}-space padding 
for the iterative gradient descent when memory allowed and
made use of GPU acceleration, and needed to vary the number of alignment iterations for parallax
to achieve optimal results.
In general, aberrations needed to be fit up to the fifth
order -- we noticed that third-order software correction on a fifth-order 
hardware-corrected instrument often resulted in worse phase uniformity than
using no aberration correction. For some reconstructions, fitting 
seemingly did not work very well in this semi-atomated workflow, 
as phase uniformity did not improve or even worsened.
\begin{table}[b!]
	\caption{Data sizes (in megabytes) and computational timings (in seconds) for
		different phase reconstruction algorithms for the full 512$\times$512
		real-space scan positions (\emph{\textbf{R}}px) with different degrees of
		reciprocal-space~binning resulting in different numbers of
		pixels (\textbf{\emph{Q}}px) in the convergent-beam electron diffraction
		patterns (CBED). The listed algorithms are integrated center of mass (iCOM), 
		parallax-corrected bright-field (parallax), iterative
		gradient descent (iter. GD), single-sideband (SSB) and Wigner
		distribution deconvolution without and with aberration fitting~(AC).
		{\label{tab:data}}%
	}
\begin{tabular}{@{\extracolsep\fill}lllllllllllllll@{}}\toprule
Scan (\textbf{\emph{R}}px) & Bin in \textbf{\emph{Q}} & CBED (\textbf{\emph{Q}}px)
 & Array (MB) & iCOM (s) & Parallax (s) & Iter. GD (s) & SSB (s) & SSB~(AC) (s) & WDD (s) & WDD~(AC) (s) \\
 \midrule
512$\times$512 & 2 & 96$\times$96 & 9664 & 10   & --     & 1198& 365 & 390 & 378 & 382\\
                             & 4 & 48$\times$48 & 2416 & 8.8 & 157   & 698 & 113 & 118 & 156 & 158 \\
                             & 8 & 24$\times$24 & 604   & 8.1 & 153   & 242  & 43 & 48 & 44 & 44 \\
                             & 16 & 12$\times$12 & 151    & 8.0 & 16     & 135  & 25 & 29 & 26  & 27 \\    
                             & 32 & 6$\times$6    & 38     & 7.6 & 19     & 108  & 20& 23 & 21 & 20 \\
 \bottomrule
\end{tabular}
\end{table}

\begin{figure}[h!]
\begin{center}
\includegraphics[width=0.6\columnwidth]{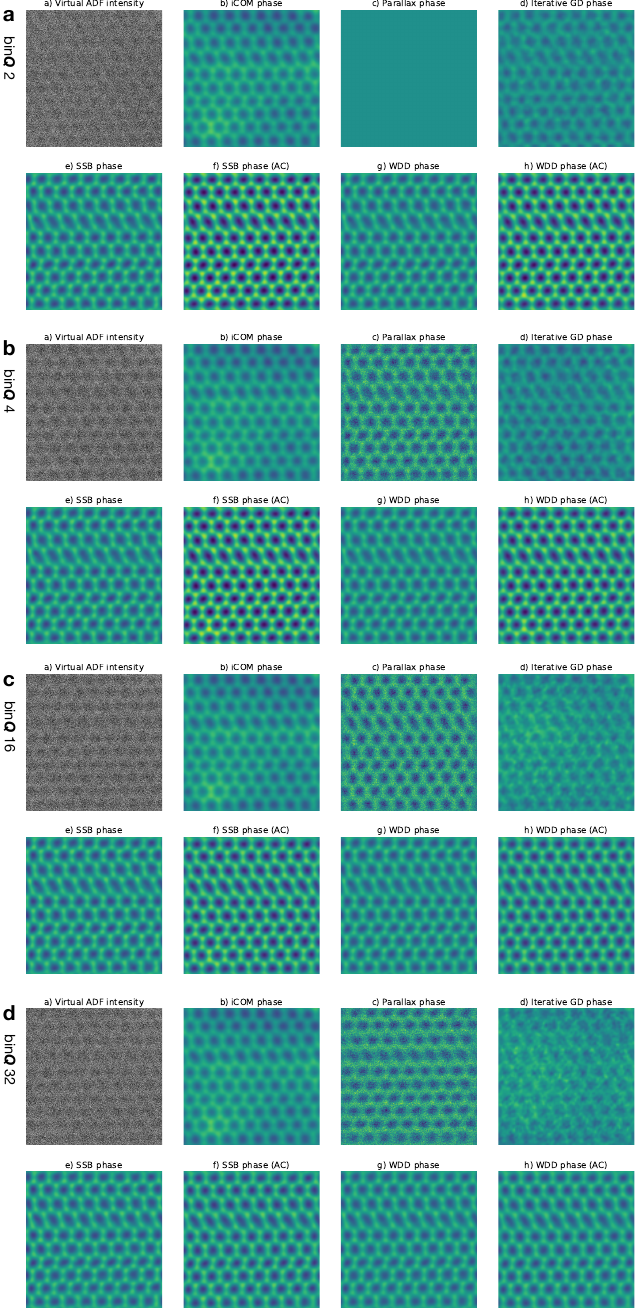}
\caption{Comparison of the effect of \textbf{\emph{Q}}-binning on phase images of the 36 mrad maximum scattering angle dataset recorded with a pixel dwell time of 33.3~\si{\us} over pristine
monolayer graphene reconstructed with the different algorithms indicated
over each panel. a) shows images for a \textbf{\emph{Q}}-binning factor of 2, b) for 4, c) for 16, 
and d) for 32. The field of view is 2.1$\times$2.1~nm\textsuperscript{2} and the phase intensity 
scale ranges from -15 to 15 mrad.
{\label{fig:full_binQ}}%
}
\end{center}
\end{figure}

Visually, the phase images look similar across the range of binning
factors considered here. Specifically, there is little change between 2$\times$
and 16$\times$ binning for this magnified setting that is optimal for bright-field
ptychography except for the aberration-corrected SSB and WDD, where contrast
is noticably increased. I therefore display in
Figure~{\ref{fig:full_binQ}} only the phase images
reconstructed from the 2$\times$, 4$\times$, 16$\times$ and 32$\times$ binned data for each of
the algorithms, alongside a virtual annular dark-field (ADF) image with
the inner radius set just outside the bright-field disk. Such a large magnification of the
diffraction patterns has the benefit of maximizing the collected bright-field signal,
but is obviously disadvantageous for virtual ADF imaging. Notably, a
distortion likely due to small stage jump is visible as a vertical
``stretching'' about one third of the way down the 2.1$\times$2.1~nm\textsuperscript{2}
field of view. The atomic phase variation is quantified in Section~{\ref{sec:quantification}};
importantly, the phase optimization method I use is able to at least partially account for such distortions. 

Remarkably, although the parallax method nominally requires bright-field
contrast to align the pixels corresponding to different scattering
vectors, I found parameters that resulted in very good results even for
such a weakly scattering object. Important here was to allow only a few 
(between one and three, typically) alignment iterations at the largest possible
binning (typically 32). 
Further, while the iterative gradient descent algorithm is typically intended for defocused-probe
reconstructions, with a judicious selection of parameters also it was
also able to perform rather well here albeit at a lower resolution due
to the limited~\emph{\textbf{Q}-}space padding allowed by memory -- 
as we will see below, in some situations it may even outperform other 
algorithms.

Regarding phase uniformity, it is clear that iCOM, and to a smaller extent iterative gradient descent, 
exhibit greater low-frequency phase variation than the other methods. The signal-to-noise ratio in 
iCOM is known to suffer at small spatial frequencies \textbf{\emph{k}}, with an analytical expression 
inversely proportional to \textbf{\emph{k}}~\cite{varnavides_iterative_2024}. For iterative GD, the 
reason is analytically less clear, but it appears empirically that small spatial frequencies are the 
hardest to converge, and the exclusion of larger scattering angles by the magnified projector lens 
seitting is further disadvantageous for this algorithm. Note also that although it would easily improve 
phase uniformity, for unbiased comparisons I did not here apply any smoothing 
or high-pass filtering (which would be directly supported within 
the reconstruction algorithms in py4DSTEM).

\subsubsection{Fastest acquisition hardware-binned data}

\begin{figure}[h!]
	\begin{center}
		\includegraphics[width=0.6\columnwidth]{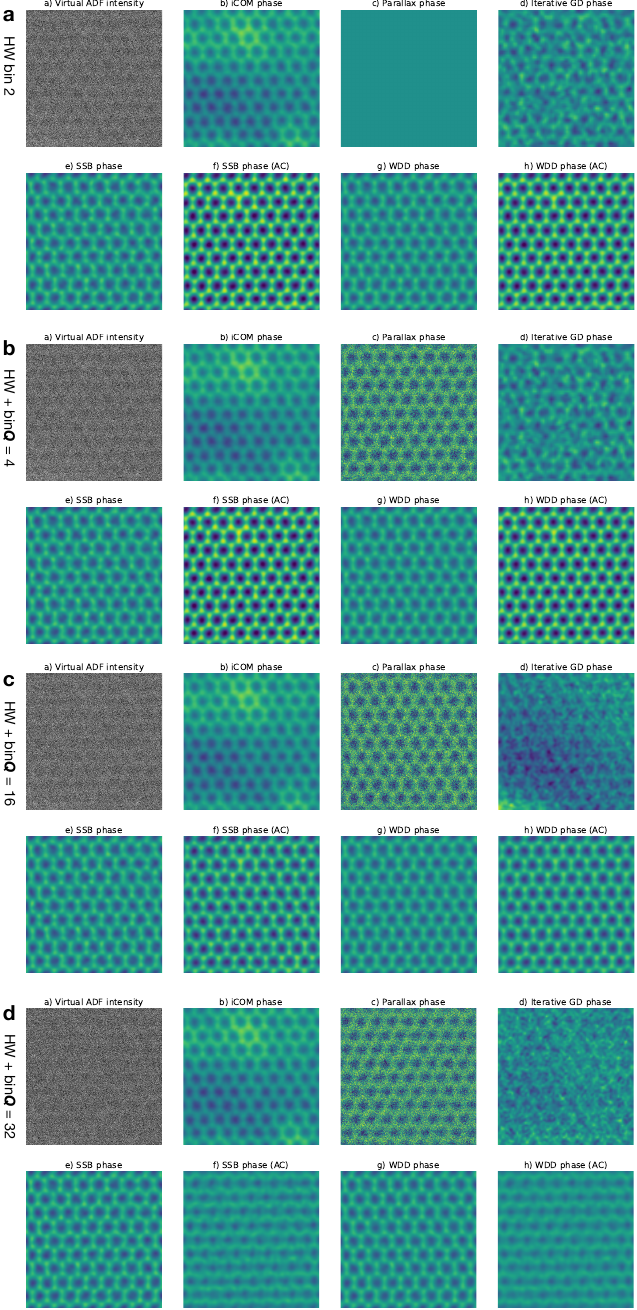}
		\caption{Comparison of the effect of \textbf{\emph{Q}}-binning on phase images of the 36 
		mrad maximum scattering angle dataset recorded with pixel dwell time of 10.0~\si{\us} over 
		pristine monolayer graphene reconstructed with the different algorithms indicated over each 
		panel. a) shows images for only 2$\times$ hardware binning, b) for a further software 
		\textbf{\emph{Q}}-binning factor of 2 (total 4), c) for 8 (total 16), and d) for 16 (total 32). The 
		field of view is 2.1$\times$2.1~nm\textsuperscript{2} and the phase intensity scale ranges 
		from -15 to 15 mrad.
			{\label{fig:fast_binQ}}%
		}
	\end{center}
\end{figure}

With such a large bright-field disk recorded on the camera, significant binning
is possible and even desirable. Before I discuss real-space thinning, let us thus 
first consider a separate dataset collected with the 2$\times$ hardware-binning required 
to run the ARINA detector at its maximum frame rate of 120,000 fps. Although the 
beam current was still very high at 207 pA, the shorter pixel dwell time resulted in 
three times lower doses than in the full dataset. Again, the phase images look similar 
across the range of binning factors, so I display in Figure~{\ref{fig:fast_binQ}} only the phase 
images reconstructed from only the hardware binning, and additional 2$\times$, 8$\times$ and 
16$\times$ software-binned data for each of the algorithms. The distortion due to the stage shift 
seen in the previous dataset is notably not present here.

Although broadly speaking the results are similar to the full data in Figure~{\ref{fig:full_binQ}},
with rather high additional binning still being possible, some interesting differences can be 
discerned. The iterative gradient descent method cleraly struggles with this data, presumably due 
to the lower dose that was collected, but parallax remains robust throughout. Remarkably, 
although aberration correction becomes poor at the highest binning, naive SSB and WDD retain 
atomic resolution even for the very highest total binning factor of 32, which corresponds to 
diffraction patterns with only 6$\times$6 pixels. Clearly data sizes can be significantly reduced by 
binning, as has also been noted before, but such a large projector lens magnification optimizes 
this further. It is unforunate the ARINA is not able to bin by more than a factor of two in hardware,
since greater binning would further reduce the data rate and help unleash the true maximum 
speed of the underlying electron counting application-specific integrated circuit of the 
detector~\cite{zambon_kite_2023}.

\subsubsection{Magnified projector lens setting: software thinning in~\textbf{\emph{R}}}
Before considering the extended projected lens setting, let us first investigate
the effect of thinning the full data in~\textbf{\emph{R}}, that is, omitting
every~\emph{n}th scan pixel, which both reduces the sampling and the
total dose available for the reconstructions. The real-space sampling of the
original data was 0.041~\AA/pixel, corresponding to a relative probe overlap 
(defined as the linear distance offset) of 96\% (assuming a round probe with full-width at half-maximum of 1.1~\AA), which reduces to 70\% for a thinning factor of 8
(real-space sampling of 0.33 \AA/pixel), just above the criterion proposed by Bunk~\cite{bunk_influence_2008}.
For the comparisons shown in Figure~{\ref{fig:full_thinR}, I again used the full 
192$\times$192 \textbf{\emph{Q}}px data acquired with a dwell time of 33.3~\si{\us} 
per pixel with a \textbf{\emph{Q}}-binning factor of 16 (note that for ease of comparison,
the top-most images labeled (a) thus reproduce the images labeled (c) in~Figure~{\ref{fig:full_binQ}}). 
The difference to \textbf{\emph{Q}}-binning is striking for iCOM, with atomic contrast being 
entirely lost for thinning factor 8. However, the other algorithms are still able to retain good 
resolution, with iterative gradient descent even benefitting from the reduced sampling.

\begin{figure}[h!]
	\begin{center}
		\includegraphics[width=0.6\columnwidth]{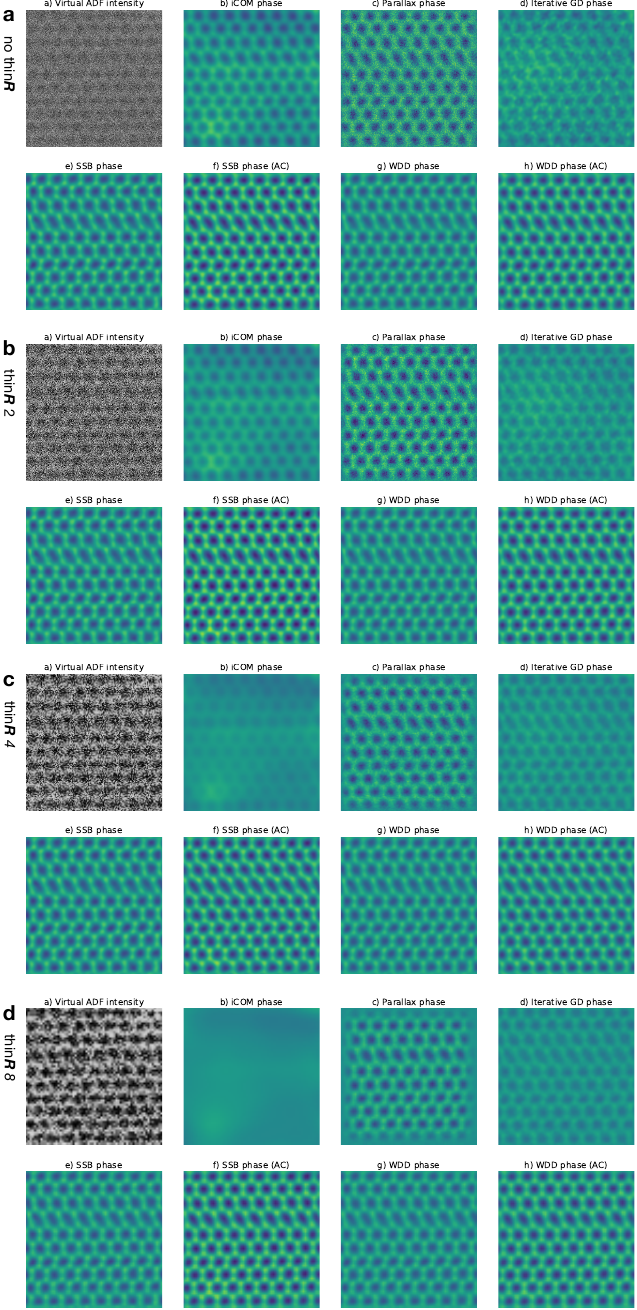}
		\caption{Comparison of the effect of \textbf{\emph{R}}-thinning on phase images of the 36 mrad maximum scattering angle dataset recorded over pristine
				monolayer graphene reconstructed with the different algorithms indicated
				over each panel. A \textbf{\emph{Q}}~binning factor of 16 was used throughout. a) shows images for
				no thinning, b) for a \textbf{\emph{R}}~thinning factor of 2, c) for
				4, and d) for 8. The field of view is 2.1$\times$2.1~nm\textsuperscript{2} and the phase intensity scale ranges from -15 to 15 mrad.
				{\label{fig:full_thinR}}%
		}
	\end{center}
\end{figure}

\subsubsection{Extended projector lens setting: software binning in~\textbf{\emph{Q}}}
Phase reconstructions of the data recorded with the extended projector lens setting with 109 mrad maximum recorded scattering angle with 
256$\times$256 real-space~\textbf{\emph{R}} scan positions with different
degrees of reciprocal-space~\textbf{\emph{Q}} binning are compared in Figure~\ref{fig:long_binQ}.
Although the overall visual impression is similar to the magnified setting images
despite this data including fewer scan positions (corresponding to thinning factor 2 with respect
to the magnified data), a few differences can be easily noted. Unsurprisingly, the virtual ADF images 
have much better contrast now that larger scattering angles are included. Iterative gradient descent 
reconstructions now show a visibly greater spatial resolution for most binnings, and parallax 
continues to perform well.
The direct ptychography algorithms also do not fare too poorly enve when sampling of the bright-field
disk decreases to below 8 pixels for the binning factor of 8, although fitting aberrations correctly
clearly becomes increasingly difficult. However, naive SSB  and WDD still perform remarkably
well, suggesting extended projector lens settings with moderate binning may provide a good compromise
when good virtual ADF imaging is desirable and both direct and iterative methods are used.
However, some differences in the phase variation can be noted, as will be discussed next in
section~\ref{sec:quantification}.

\begin{figure}[h!]
	\begin{center}
		\includegraphics[width=0.6\columnwidth]{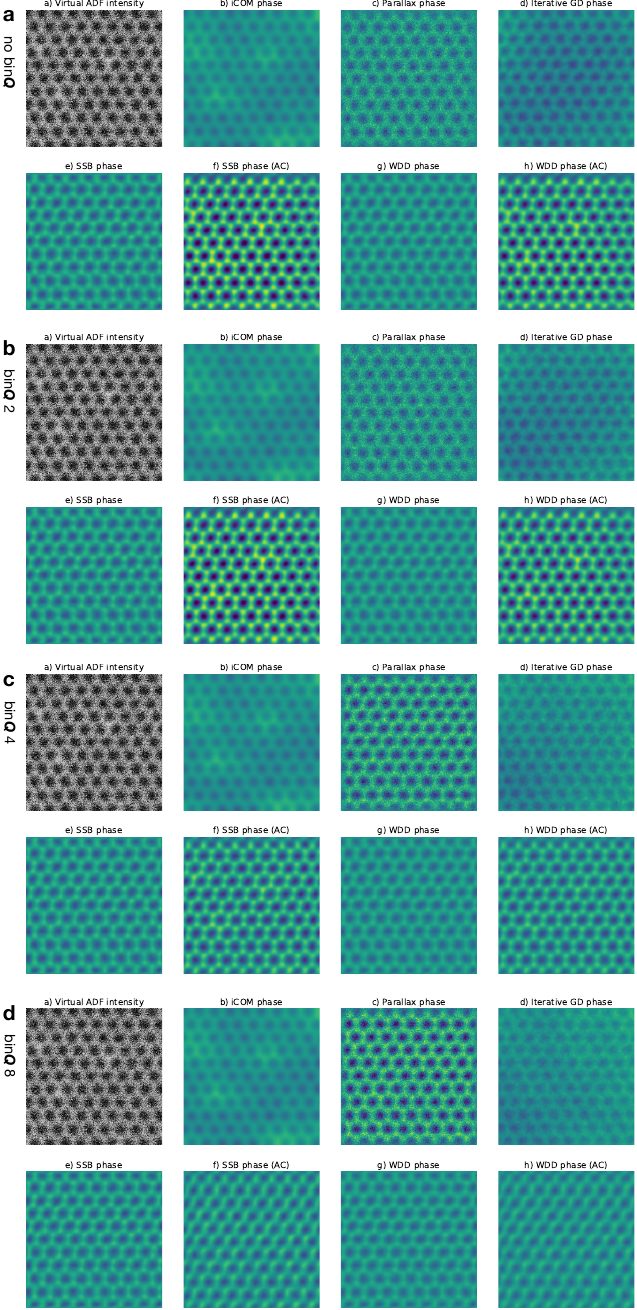}
		\caption{Comparison of the effect of \textbf{\emph{Q}}-binning on phase images of the 109~mrad maximum scattering angle
			dataset recorded with a pixel dwell time of 100~\si{\us} over pristine
			monolayer graphene reconstructed with the different algorithms indicated
			over each panel. a) shows images for no binning, b)
			for a \textbf{\emph{Q}}-binning factor of 2, c) for 4, and d) for
			8. The field of view is 2.1$\times$2.1~nm\textsuperscript{2} and the phase intensity scale ranges from -15 to 15 mrad.
			{\label{fig:long_binQ}}%
		}
	\end{center}
\end{figure}

\subsection{Quantification of phase variation}
{\label{sec:quantification}}
Finally, let us turn to the quantification of the atomic phase shifts from the
reconstructed phase images. Due to the complicated nature of the 
contrast transfer function of SSB~\cite{yang_efficient_2015} (specifically, atomic phase contrast has a negative halo which can influence the
phase at neighboring sites~\cite{hofer_reliable_2023}) and other ptychography methods~\cite{oleary_contrast_2021}, and due to the effect of scan
distortions, drift, or sample tilt, simple methods based on e.g. Gaussian fitting
or Voronoi integration may not give reliable results. I therefore used the
parameter-based iterative method developed by Hofer and Pennycook~\cite{hofer_reliable_2023}
as described in the Methods. 

Briefly, in this optimization method, an initial atomic model is created to
correspond to the visible part of the lattice. This is then converted to a
point potential based on the model, after which the contrast transfer
function (CTF) of a given ptychography method is converted to the point-spread
function applied to each position, resulting in a phase image matching the model positions and
intensities. The model is then iteratively optimized so that the correlation
between the model image and the experimental phase image is maximized.
Notably, this optimizes not only the positions of the atoms in the model, but
also the strengths of the point potential at each location, from which I then
derive the atomic phase shifts -- and, crucially for this study -- their variation
over the lattice. The means of the distributions of atomic phase values 
estimate the absolute phase magnitude for each method, while their variation
estimates how well each algorithm copes with noise and aberrations.

As an improvement of the original phase optimization method~\cite{hofer_reliable_2023}, the CTFs of
iCOM, parallax and iterative gradient descent method~\cite{varnavides_iterative_2024} are also now explicitly included. One caveat should however be noted: the convergence of the method currently requires
the handcrafting of a relatively accurate model for the field of view, which
changes not only between acquisitions, but may vary slightly depending on
the exact reconstruction algorithm. This led to the fitting occasionally failing,
necessitating a tedius manual verification and modification of the model. Clearly, a better
approach in the future will be to use machine vision to detect the lattice 
and to create the model automatically to match each phase image.

Figure~\ref{fig:optimization} shows examples of the procedure for the full 
magnified projector lens setting dataset with a \textbf{\emph{Q}}-binning factor of 16. This 
is a particularly challenging case due to the scan distortion, but as we can see,
 the iterative optimization procedure is able to satisfactorily though not
perfectly account for the imperfections of the shown SSB image (panel a). For parallax, the 
experimental phase images contained much more shot noise that led to
spuriously low correlations, so these were smeared with a Gaussian of width 0.25~\AA\ before 
optimization. The convergence of the correlations for each method are shown in panel b;
I note that the somewhat low values are partly explained by the edges, but the distortion does 
contribute. As contrast at the edges was not correctly reproduced, 15\% of the field of view from 
each side was always omitted from the further phase quantification.
In panel c, histograms corresponding to the absolute atomic phase magnitudes are shown
for each of the methods (with panel d showing the corresponding relative spread).
                         
\begin{figure}[h!]
	\begin{center}
		\includegraphics[width=1.0\columnwidth]{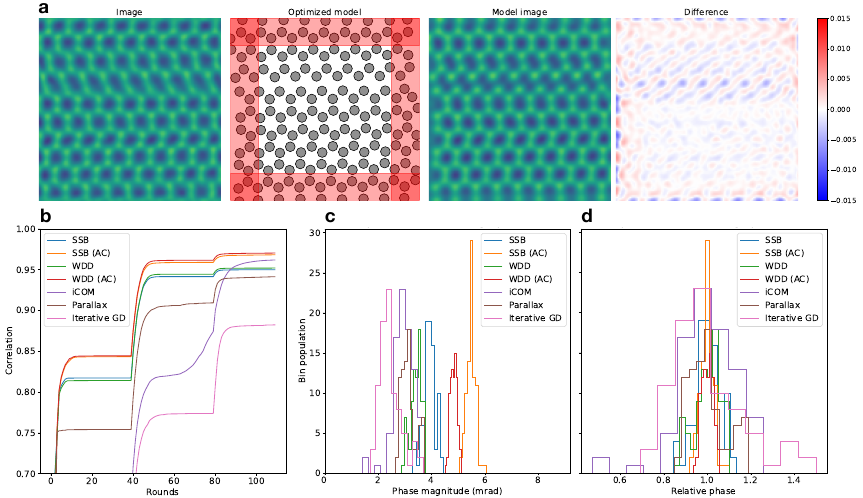}
		\caption{Example of atomic phase quantification for the full magnified projector lens setting
			dataset with a \textbf{\emph{Q}}-binning factor of 16. a) The experimental
			SSB phase image, the optimized atomic model matching the field of view	after a three-stage 
			optimization procedure with 110 rounds in total (see Methods), the simulated model image, 
			and its difference to the experimental one (colormap 'bwr'). The phase intensity color scale 
			spans from -15 to 15 mrad. b) Correlation between experimental and optimized model images 
			as a function of iteration rounds for all reconstruction algorithms. c-d) Distributions of the 
			absolute (c) and relative (d) optimized atomic phase shifts (with atoms within 15\% of the edge 
			excluded as shown in red on the atomic model in panel a) for all reconstruction algorithms.
			{\label{fig:optimization}}%
		}
	\end{center}
\end{figure}

The quantication is presented in Figure~\ref{fig:quantification} for all binning and thinning values 
including the few that were omitted from the images shown in 
Figures~\ref{fig:full_binQ}--\ref{fig:long_binQ}. Overall, the absolute values of the atomic phase
shifts are in relatively good agreement, with SSB and WDD especially for the less binned data with 
fitted aberration correction showing slightly larger values. In terms of the phase variation, iCOM (not 
to mention the ADF images that were not quantified) and parallax appear to perform somewhat more 
poorly than SSB or WDD, whereas iterative gradient descent produces 
slightly greater variation with the magnified projector lens setting and especially for the faster dataset, but performs 
quite well in comparison for the data thinned in \textbf{\emph{R}} as well as the extended setting. 
Comparison of the full and fast magnified setting data for SSB and WDD with a 
threefold difference in dose suggests
that these reconstructions are not fluence-limited, as the variation of phase is even smaller for the lower-dose data.
The unexpected larger variation of the aberration-corrected full data where a sample shift was 
present (visually apparent in the top panels of Figure~\ref{fig:full_binQ}) may indicate that 
this distortion has interefered with the accurate fitting of the abberation coefficients.

\begin{figure}[h!]
	\begin{center}
		\includegraphics[width=1.0\columnwidth]{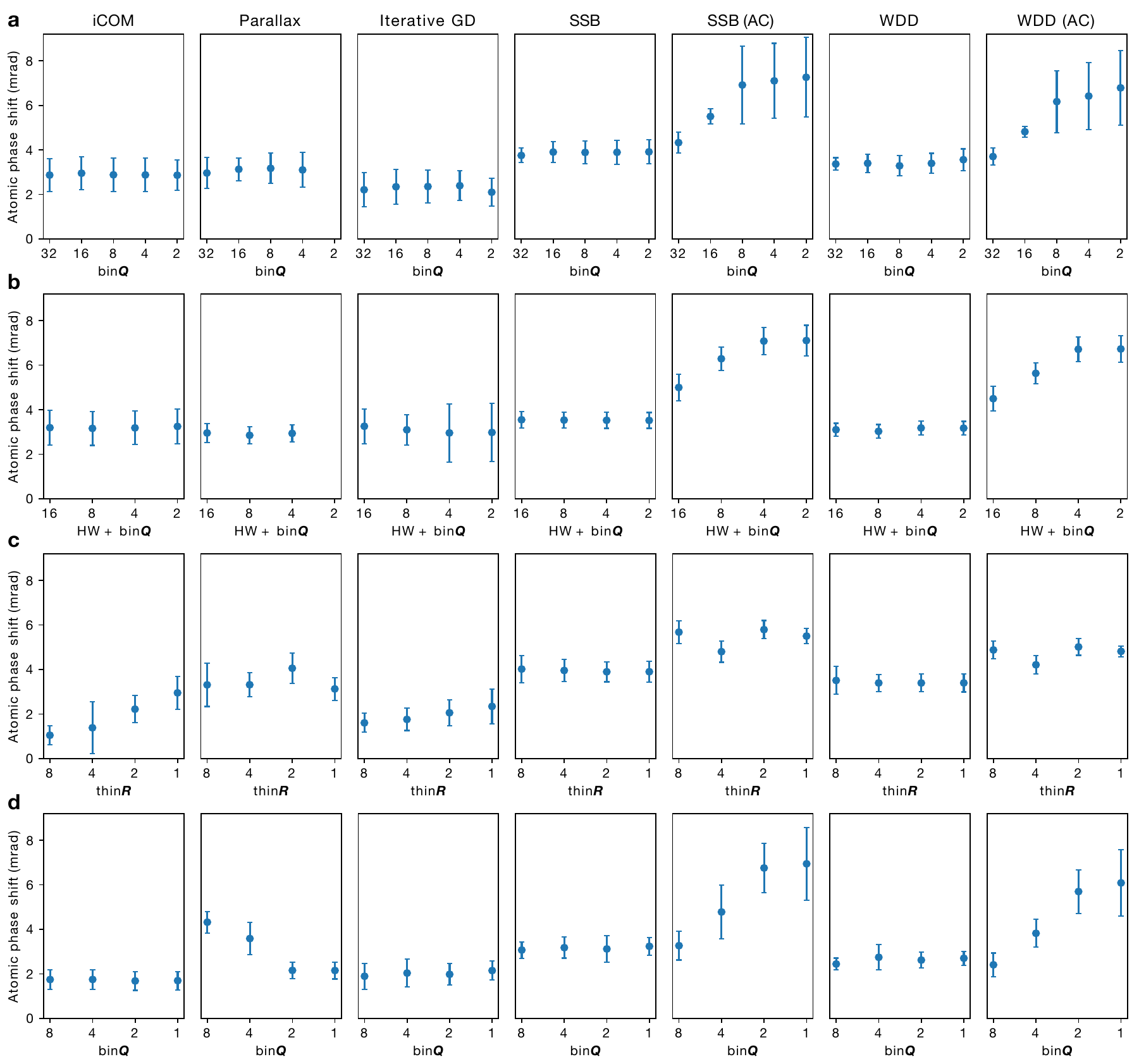}
		\caption{Quantification of the atomic phases for each of the reconstruction
			algorithms for different datasets and levels of binning and thinning. The points show 
			mean phases with 95\% confidence intervals based on their standard error.
			a) Magnified projector lens setting full dataset \textbf{\emph{Q}}-binning. b) Magnified projector lens setting fast dataset \textbf{\emph{Q}}-binning.
			c) Magnified projector lens setting full dataset \textbf{\emph{R}}-thinning. d) Extended projector lens setting \textbf{\emph{Q}}-binning.             
			{\label{fig:quantification}}%
		}
	\end{center}
\end{figure}

\subsubsection{Dependence of WDD phase on $\epsilon$}
Finally, the dependence of the WDD phase values on the Wiener filtering parameter $\epsilon$ requires a separate discussion. The original theory~\cite{rodenburg_theory_1992} and its 
algorithmic implementations~\cite{hofer_pyptychostem_2024}
include an additive parameter $\epsilon$ in the denominator of the Wigner distribution term.
This is meant to act as a Wiener filter, and its value should be small and presumably is not meant to
influence the results significantly. I originally found that a suggested default value of 0.01 
produced poor-quality phase images, and thus settled on the smallest value that did at 0.05. 
However, upon further inspection it turned out that the WDD phase magnitudes converge towards
the SSB values when $\epsilon$ \emph{increases} toward 1.

To quantify this, I reconstructed selected datasets (using iteratively fitted aberration coefficients,
though the results are similar without) with different values of $\epsilon$ and, for simplicity and to 
show this effect is not dependent on the CTF-based optimization, compared the maximum phase 
values to SSB reconstructions of the same data. The 
results are shown in Figure~\ref{fig:wdd_variation}. The convergence of the maximum 
phase values as $\epsilon$ increases is clearly apparent, though there still seems to be small
discrepancy compared to the respective SSB phase values. 
Such large values of $\epsilon$ appear to go against the original intention behind the 
algorithm, suggesting 
that the original theoretical derivation may need revisiting due to the effect the Wiener
filter has on realistic experimental data (please also see Section~\ref{subsec:methods:reconstruction} 
for discussion of replication of the findings with an independent implementation).

\begin{figure}[h!]
	\begin{center}
		\includegraphics[width=0.5\columnwidth]{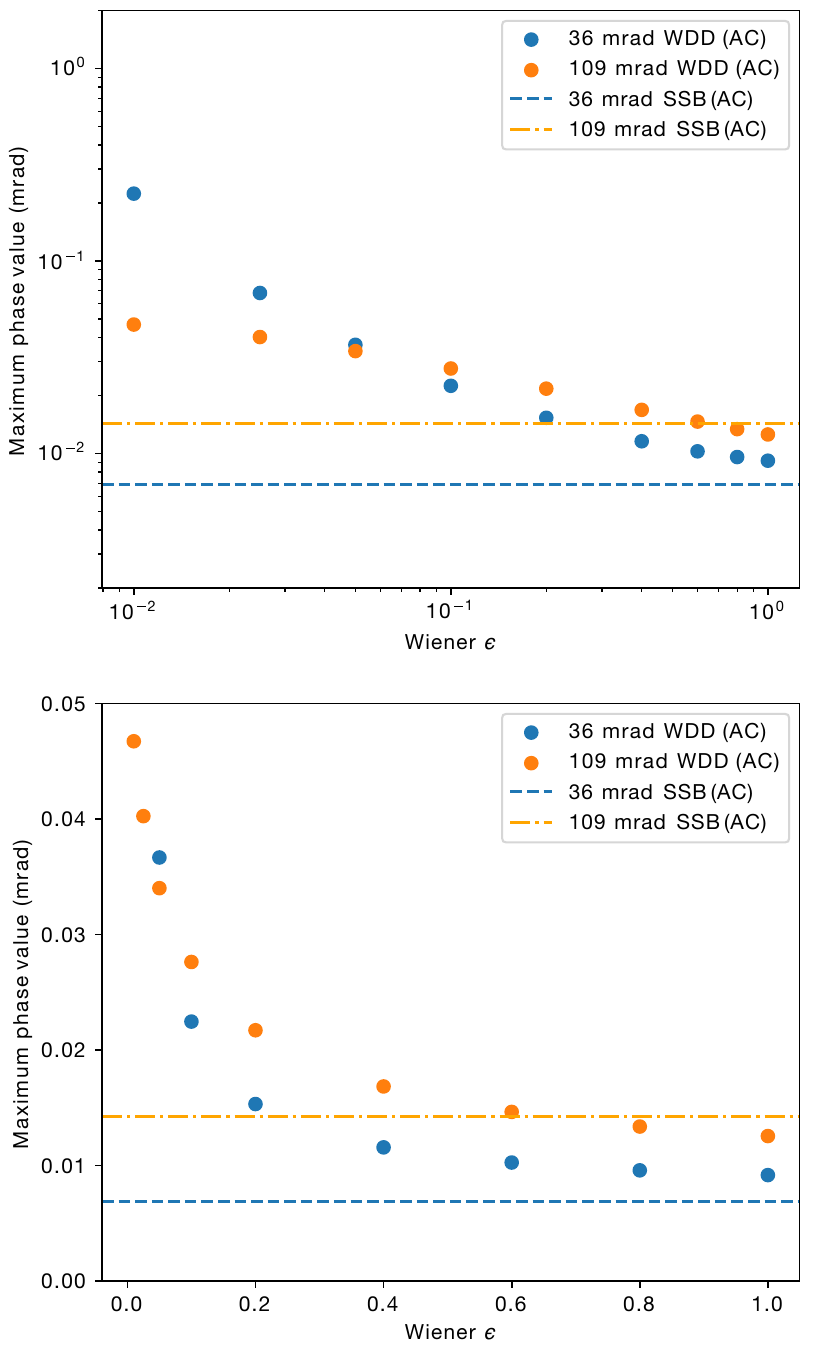}
		\caption{Convergence of the aberration-corrected WDD maximum phase value 
			as a function of increasing Wiener $\epsilon$ parameter compared to respective
			SSB value for two datasets: full magnified projector lens setting data binned by 4 in 
			\textbf{\emph{Q}}, and the unbinned extended projector lens setting data.
			a) A log-log scale plot showing the full variation. b) A linear scale with the largest WDD phase 
			values at small $\epsilon$ cropped.
			{\label{fig:wdd_variation}}%
		}
	\end{center}
\end{figure}

\section{Conclusions}
To summarize, I have used graphene as an ideal uniform phase object to
compare direct and iterative electron-ptychographic phase reconstruction
algorithms with data collected at high electron dose in an effort to benchmark
the algorithms' performance and reliability. The electron optics were aligned
 with a maximally magnified projector lens setting for the
 optimal sampling of the bright-field 
disk, and I compare that to an extended projector lens setting typically used for defocused-probe 
data collection and super-resolution. I have evaluated the algorithms' computational 
efficiency and tolerance to reciprocal-space binning and real-space thinning of the data,
and quantified the atomic phase shifts to provide guidance for the sampling required for
desired levels of relative phase precision with each method. Although the methods mostly agree,
iterative gradient descent clearly benefits from collecting larger scattering angles,
whereas direct methods are faster and can operate reliably with sampling only 
a few dozen pixels of the bright-field disk. My findings also suggest that the
Wigner distribution deconvolution phase values sensitively depend on the Wiener
filter parameter in a way that appears to go against the original intention of the algorithm. 
By using exclusively open-source software tools and sharing my data and 
code openly, I hope this contribution helps spur further development and standardization of
electron ptychography as a truly quantitative technique for the detection of 
electrostatic potentials, charge transfer, magnetism, and beyond.

\section{Methods}

\subsection{Sample preparation}

Commercial monolayer graphene grown by chemical vapor deposition and
transferred by the vendor onto TEM grids with Quantifoil carbon foil
supported on an Au mesh from Graphenea were used. Clean and defect-free
(pristine) monolayer regions were identified by imaging and the 4D-STEM
data subsequently recorded. The scanned area was set
to approximately 2.1$\times$2.1 nm\textsuperscript{2}, as with the lattice 
constant of graphene this conveniently produced a nearly periodic field of view.

\subsection{4D-STEM data acquisition}

Data was recorded using the Nion Swift software~\cite{meyer_nion_2019}
(version 0.16.10), acquired on the Nion UltraSTEM 100
aberration-corrected STEM instrument operated at 60~keV with a probe
convergence semi-angle of 34 mrad as verified from the averaged
convergent-beam electron diffraction pattern and instrument parameters,
and the Dectris ARINA direct-electron detector (with Si sensor
material). The beam current was close to 200~pA for all collected
datasets, and the sample was at room temperature in ultra-high vacuum
(near 10\textsuperscript{--9}~mbar).

Different magnifications were set up using the projector lens system of
the microscope, which I denoted by the maximum scattering angle
captured by the detector array in the vertical and horizontal directions
(diagonal scattering angles were correspondingly larger), namely 36~mrad
($\sim$$\alpha$) and 109~mrad ($\sim$3$\alpha$). This allowed us
to study the effect of the experimental projector lens setting on the
reconstruction performance, ranging from maximal sampling of the
bright-field disk to a setting more conducive to iterative algorithms
and super-resolution images (the latter of which was explictly not my
aim here).                                                           

Most data was collected using the full unbinned 192$\times$192 pixel
(\textbf{\emph{Q}}px) array and binned to varying degrees in software to
study the tolerance of the algorithms to limited reciprocal-space
sampling, with a pixel dwell time of 33.3~\si{\us}, but some data used 2$\times$
hardware binning to 96$\times$96 pixels to reach the highest possible frame
rate with pixel dwell time of 10~\si{\us}. For the magnified projector lens setting scan,
I used 512$\times$512 pixels for sufficiently high sampling, which was thinned     
in software by omitting scan positions to study the effect of limited
real-space sampling. For the extended projector lens setting, I used fewer 256$\times$256 scan
pixels to keep the total acquisition time roughly constant with the longer 100~\si{\us} pixel dwell 
time used to increase signal for larger scattering angles. For the purposes of the iterative gradient 
descent reconstructions, the probe was separately recorded over vacuum for each
setting.
       
\subsection{Phase reconstructions}
\label{subsec:methods:reconstruction}
All ptychographic phase reconstructions were performed using the py4DSTEM 
package~\cite{savitzky_py4dstem_2021,varnavides_iterative_2024} (version~0.14.19). For SSB 
and WDD, apart from the experimental settings -- probe convergence semi-angle, primary beam  
energy, scan-step size, and \textbf{\emph{Q}}--\textbf{\emph{R}} rotation (see below) -- for WDD 
the only other adjusted 
input parameter was
the small positive constant $\epsilon$ introduced to avoid dividing by
zero values in the Wigner deconvolution. I initially used a value of 0.05 for this constant, but 
as discussed in Section~\ref{sec:quantification}, I found that counter-intuitively,
the larger this value, the smaller were the absolute phase magnitudes and the
more in line with the other methods they were. I therefore settled on a value of
to 1.0 to achieve converged phase values for the results presented here.

When aberration correction was enabled for SSB and WDD, the aberration
coefficients were recursively fitted up to 5th radial order based on five sets of
double-disk overlaps. However, it should be noted that not doing aberration
correction merely assumes the values of all coefficients to be zero without 
providing any advantage in terms of simulation time. In early stages of this work presented at conferences~\cite{susi_opensource_2024,susi_ideal_2024}, I have used the
PyPtychoSTEM package~\cite{hofer_pyptychostem_2024} instead, which 
produced similar results. PyPtychoSTEM did require somewhat less memory but
produced less reliable reconstructions especially when fitting aberrations, though the computational
speed of SSB without aberration correction was also significantly faster. Notably, the
same sensitivity of the WDD phase magnitudes to the value of $\epsilon$ was present with
that code. This suggest to me that this is not an issue with the implementations, but rather 
an unexpected and frankly undesirable feature of the original theory, at least
when applied to realistic electron-ptychographic data.

For the iterative reconstructions, the py4DSTEM iterative differential
phase contrast method (essentially identical to integrated center of mass,
which is how I label it here to avoid confusion) 
was used to verify the rotation angle between the scan
and the reciprocal-space directions (\textbf{\emph{Q}}-\textbf{\emph{R}}
rotation) determined by the camera orientation -- these were set to be
as close to aligned as possible when the electron-optical setup was made
-- resulting in an angle of typically 3 degrees found by minimizing the
curl of the $x$-$y$ center-of-mass (COM) signals. Notably, I found that
blurring these with a Gaussian of 4 px greatly improved the robustness
of finding the angle, and that 20 iterations of the algorithm was more
than sufficient for convergence. Importantly, I uncovered a factor of
2$\pi$ difference between this and the other ptychography methods,
which I have corrected by hand in all of my analyses (pending code revision
by the py4DSTEM developers).

For parallax-corrected bright-field imaging, an intensity threshold of
0.65 for detecting the bright-field disk pixels was used, with the 
\textbf{\emph{Q}}--\textbf{\emph{R}} rotation
forced to the previously determined value. Only the coarsest possible
binning (bin value of 32, or the closest to it if the data was binned
more) and between 1 and 5 iterations was used to align the different
tilts; crucially, I found that any further iterations or finer binning
resulted in poor subpixel alignment and contrast transfer
function fitting, and thus resulting phase images, presumably due to the
extremely low bright-field signal level from a lattice consisting of a
single layer of light elements.

Finally, for batched iterative gradient descent single-slice
ptychography, the~\textbf{\emph{Q}}--\textbf{\emph{R}} rotation was again
forced, the object type was set to `potential' and object positivity
set to False (to treat the specimen potential as a pure weak phase
object), and the probe was allowed to update (as fixing it did not result in good
reconstruction results). The object was padded by 12 pixels in both directions, and the
diffraction-space padded by a factor of 3 or 2 times whenever memory allowed to
improve the resolution of the reconstruction for the magnified projector lens setting. Separately
recorded vacuum probes were provided to the algorithm to reduce
probe-object mixing. The batch size was set to 512 for optimal memory use, and 15                         
iterations were found to be sufficient for convergence, with more resulting in more severe
mixing in my testing. Other recontruction parameters were at the default
settings, including the update step size of 0.5 and a weight of the maximum 
probe overlap intensity of 1. More information can be found in Ref.~\cite{varnavides_iterative_2024}.

All reconstructions were performed on a 32-Core AMD Ryzen 
Threadripper 1950X (parallelized over workers to different degrees depending on the
algorithm), except for the batched iterative gradient descent, which was
accelerated using CuPy on an NVidia RTX4090 GPU (with data stored in the
CPU RAM so that sufficient memory was available for all datasets). Parallax
would also benefit greatly from GPU acceleration, but memory usage became
a bottleneck as CPU RAM data storage was not available there. However, the relative performance
of the algorithms may be different on hardware with different constraints
(single-core performance vs. number of cores, memory size vs. memory 
bandwidth, and so on). Further, it needs to be acknowledged that these timings
represent a specific implementation of these algorithms on specific hardware, and so
broader conclusions on the inherent limitations of the algorithms themselves must be
drawn only with care.

\subsection{Phase quantification}

The atomic phase shifts were determined by parameter-based iterative
optimization~\cite{hofer_reliable_2023} (commit 
8860c37e of the 'newctfs' branch of the forked repository at the author's 
GitLab~\cite{hofer_stem_optimization_2025}).
The optimizations were based on an atomic model, which was a 9$\times$5 orthogonal                                                                                                                                                
supercell of graphene that was manually translated, wrapped and
occasionally cropped and scaled to roughly match the fields of view of
each dataset. Each experimental phase image was binned by 4~for speed
(except for the iterative gradient descent images, which had a lower
resolution, and the data thinned in \textbf{\emph{R}} which had fewer pixels), and the model was iteratively modified to maximize the
correlation between a phase image simulated with the contrast transfer
functions of each ptychography method with scattering potentials placed at the atomic
positions of the model and the experimental images. Optimization proceeded in three stages: 40 rounds of quick
optimization matched the field of view, translation and scale (and 
blur, which was not used), then 40 further rounds matched also the
positions of the model atoms, and finally 20--80 rounds optimized
also their potential strengths. These then yielded the atomic potential strengths
corresponding to the atomic phase shifts, which were binned into                scistograms to quantify the phase variation after atoms within 15\% of
the edges of the field of view were omitted. The mean (based on the bin centers) 
and standard error of the mean (at 95\% confidence level) of the phase variation were
additionally calculated for plotting.                   

\section{Data and Code}

{The 4D-STEM datasets stored as calibrated py4DSTEM DataCube objects used in this article are openly available on the
University of Vienna Phaidra repository
at~\url{https://doi.org/ 10.25365/phaidra.564}. All of the
analysis code~is openly available on the author's GitHub repository
at~\url{https://github.com/TomaSusi/arina-ptycho}. All of software
packages used in the analyses are open-source Python codes that are
available on their respective GitHub or GitLab pages.}

\section{Acknowledgements}

Many helpful discussions with and assistance by Georgios Varnavides, Stephanie Ribet, and Colin 
Ophus, as well as Christoph Hofer and Timothy Pennycook, are gratefully acknowledged. I am further 
indebted to Georgios for
 early access to and assistance with direct ptychography tools developed for py4DSTEM. I thank 
 Scott Findlay for proposing there was a phase discrepancy between the iCOM and the other 
 methods and for prompting me to implement additional CTFs for the phase optimization, and Colum 
 O'Leary for elucidating my questions about the WDD Wiener filter parameter. On the experimental 
 side, I 
 am thankful to Clemens Mangler as well as Russ Hayner and Andreas Mittelberger, 
and the rest of the team, for installing the Dectris detector on the Nion instrument and setting up 
both the electron-optical  
alignments and software aspects necessary for the data collection, and to Daniel Stroppa 
for providing useful insight into the detailed workings of the detector. 
Preliminary results and related analyses were presented at the Microscopy \& Microanalysis 2024~\cite{susi_opensource_2024}, European Microscopy Congress 2024~\cite{susi_ideal_2024}, and the Hong Kong Advanced Transmission Electron Microscopy 2024 conferences, where I had further useful discussions with several attendees.
However, responsibility for any possible mistakes in the analyses lies solely with the 
author.                                                                                                                                                        
      
\providecommand{\url}[1]{\texttt{#1}}
\providecommand{\urlprefix}{}
\providecommand{\foreignlanguage}[2]{#2}
\providecommand{\Capitalize}[1]{\uppercase{#1}}
\providecommand{\capitalize}[1]{\expandafter\Capitalize#1}
\providecommand{\bibliographycite}[1]{\cite{#1}}
\providecommand{\bbland}{and}
\providecommand{\bblchap}{chap.}
\providecommand{\bblchapter}{chapter}
\providecommand{\bbletal}{et~al.}
\providecommand{\bbleditors}{editors}
\providecommand{\bbled}{ed. by }
\providecommand{\bbleds}{ed. by }
\providecommand{\bbleditor}{editor}
\providecommand{\bbledition}{edition}
\providecommand{\bbledn}{ed.}
\providecommand{\bbleidp}{page}
\providecommand{\bbleidpp}{pages}
\providecommand{\bblerratum}{erratum}
\providecommand{\bblin}{in}
\providecommand{\bblmthesis}{Master's thesis}
\providecommand{\bblno}{no.}
\providecommand{\bblnumber}{number}
\providecommand{\bblof}{}
\providecommand{\bblpage}{page}
\providecommand{\bblpages}{pages}
\providecommand{\bblp}{p}
\providecommand{\bblphdthesis}{Ph.D. thesis}
\providecommand{\bblpp}{pp}
\providecommand{\bbltechrep}{}
\providecommand{\bbltechreport}{Technical Report}
\providecommand{\bblvolume}{volume}
\providecommand{\bblvol}{vol.}
\providecommand{\bbllvol}{, vol.}
\providecommand{\bbljan}{January}
\providecommand{\bblfeb}{February}
\providecommand{\bblmar}{March}
\providecommand{\bblapr}{April}
\providecommand{\bblmay}{May}
\providecommand{\bbljun}{June}
\providecommand{\bbljul}{July}
\providecommand{\bblaug}{August}
\providecommand{\bblsep}{September}
\providecommand{\bbloct}{October}
\providecommand{\bblnov}{November}
\providecommand{\bbldec}{December}
\providecommand{\bblfirst}{First}
\providecommand{\bblfirsto}{1st}
\providecommand{\bblsecond}{Second}
\providecommand{\bblsecondo}{2nd}
\providecommand{\bblthird}{Third}
\providecommand{\bblthirdo}{3rd}
\providecommand{\bblfourth}{Fourth}
\providecommand{\bblfourtho}{4th}
\providecommand{\bblfifth}{Fifth}
\providecommand{\bblfiftho}{5th}
\providecommand{\bblst}{st}
\providecommand{\bblnd}{nd}
\providecommand{\bblrd}{rd}
\providecommand{\bblth}{th}

\end{document}